\documentclass[aps,prd,epsf,showpacs,amsmath,amssymb,graphics,twocolumn,10pt]{revtex4}
\usepackage{graphicx}
\usepackage{dcolumn}
\usepackage{bm}

\newcommand{\be}{\begin{equation}}
\newcommand{\ee}{\end{equation}}
\newcommand{\ba}{\begin{eqnarray}}
\newcommand{\ea}{\end{eqnarray}}
\newcommand{\ban}{\begin{eqnarray*}}
\newcommand{\ean}{\end{eqnarray*}}

\newcommand{\eq}[1]{(\ref{#1})}

\begin{document}

\title{Naked Singularities as Particle Accelerators II}

\author{Mandar Patil \footnote{mandarp@tifr.res.in}, Pankaj S. Joshi \footnote{psj@tifr.res.in}
and  Daniele Malafarina \footnote{daniele.malafarina@polimi.it}}

\affiliation{Tata Institute of Fundamental Research\\
Homi Bhabha Road, Mumbai 400005, India}


\begin{abstract} We generalize here our earlier results
on particle acceleration by naked singularities. We showed recently
\cite{Patil} that the naked singularities that form due to
gravitational collapse of massive stars provide a suitable
environment where particles could get accelerated and collide
at arbitrarily high center of mass energies. However, we focussed
there only on the spherically symmetric gravitational collapse
models, which were also assumed to be self-similar. In this paper,
we broaden and generalize the result to all gravitational
collapse models leading to the formation of a naked singularity
as final state of collapse, evolving from a regular initial
data, without making any prior restrictive assumptions
about the spacetime symmetries such as above. We show that when
the particles interact and collide near the Cauchy horizon,
the energy of collision in the center of mass frame will be
arbitrarily high, thus offering a window to the Planck scale
physics. We also consider the issue of various possible
physical mechanisms of generation of such very high energy
particles from the vicinity of naked singularity.
We then construct a model of gravitational collapse to
a timelike naked singularity to demonstrate the working of these
ideas, where the pressure is allowed to be negative but the
energy conditions are respected. We show that a finite amount of
mass-energy density has to be necessarily radiated away from the
vicinity of the naked singularity as the collapse evolves.
Therefore the nature of naked singularities, both at classical
and quantum level could play an important role in the process
of particle acceleration, explaining the occurrence of highly
energetic outgoing particles in the vicinity of Cauchy horizon
that participate in extreme high energy collisions.

\end{abstract}

\maketitle

\section{Introduction}

Recently, an interesting observation was made by
Banados, Silk and West
\cite{BSW},
that black holes can accelerate infalling colliding particles
to arbitrarily high energies in the center of mass frame around the
horizon of an extremal Kerr black hole, provided certain restrictive
conditions were imposed on the angular momenta of the particles.
The plausibility and implications of this phenomenon in terms
of realistic astrophysical black holes was also investigated
\cite{Berti}.
Following the BSW paper, a number of works have now
further investigated this effect in the context of an extremal
charged spinning black hole, non-extremal rotating black holes,
non-rotating charged black holes, stringy black holes, Kaluza-Klein
black holes, examining similar possibilities in each case.
It is also claimed that this may be a generic property of
axially symmetric rotating black holes in a model
independent way
\cite{Wei}.
A general explanation to this effect of
unbound acceleration has also been proposed
\cite{Zasla3}.

In our recent work
\cite{Patil},
we showed that the divergence of center of mass
energy of colliding particles is a phenomenon not only
associated with black holes, but also with naked singularities
which are the final outcome of a continued gravitational
collapse of a massive star. We considered the class of
spherically symmetric, self-similar gravitational collapse
models, which lead to the formation of a naked singularity
final state developing from a regular initial data.
We considered collision between a highly energetic outgoing
particle, emerging from a close vicinity of the naked singularity,
and traveling close to the Cauchy horizon, with an ingoing
particle. We showed that the center of mass energy of such
a collision is arbitrarily large, depending on how close
is the point of collision to the Cauchy horizon.

The limitation of our work, however, was that we
considered there only the self-similar spherically symmetric
spacetimes for the sake of simplicity and definitiveness.
These models have served as a successful theoretical
laboratory to test the possible outcomes for a complete
gravitational collapse over past number of years, and
to test the validity or otherwise of the cosmic censorship
hypothesis. There have been several numerical
as well as analytical investigations of self-similar
gravitational collapse models, for different matter fields
satisfying reasonable energy conditions, such as dust
\cite{dust},
ideal fluids with non-vanishing pressures
\cite{Ori},
massless scalar fields
\cite{scalar1, scalar2},
and such others, leading to the formation of naked
singularities in gravitational collapse from a regular
initial data.

In fact, black hole and naked singularity formation
in gravitational collapse models has been investigated in
detail in recent years (see e.g.
\cite{Joshi} and references therein for analytical results
and \cite{Rezolla} for numerical results), for a wide variety of physically
realistic and relevant non-self-similar spherically symmetric
situations as well. While most of the models investigated
are spherically symmetric, certain non-spherical collapse
scenarios have also been investigated
(see e.g. \cite{Krolak}),
where the collapse can
result either in a black hole or naked singularity
final state, depending on the nature of the initial data.
The genericity
and stability of such models,
in the presence of non-spherical perturbations has
also been considered in some detail. Thus a question that
arises naturally is, whether or not the phenomenon of
divergence of center of mass energy in a collision
of particles is a generic phenomenon, or is it merely an
artifact of various symmetries imposed on the spacetime
such as the self-similarity or spherical symmetry.
We address this issue in this paper. We show here the
divergence of center of mass energy of colliding particles
near the Cauchy horizon, without making any a priori
assumptions about the spacetime symmetries such as
those mentioned above.

This implies that the ultra-high-energy particle
collisions is a phenomenon associated with naked singularities
in a much more deeper manner, independent of various
spacetime symmetries that are at times assumed for the sake
of simplicity or definiteness. This phenomenon therefore
deserves further investigation, and may have rather interesting
physical consequences in terms of observability of
the very high energy Planck scale physics.

We first make here some clarifying remarks about naked
singularities before proceeding further. Whether or not naked
singularities would occur in the real world we live in,
is an unanswered question till this date. While these are
hypothetical astrophysical objects, there is as yet no compelling
observational evidence to confirm the existence of naked
singularities unlike their black hole counterparts. However,
considering the existence and recent emergence of very many
gravitational collapse scenarios in general relativity, where
the evolution of the collapsing matter cloud from a regular
initial data leads to a naked singularity final fate for
collapse, we may assume that the naked singularities could
occur in various physical circumstances such as the final fate of
a massive star, when it undergoes a complete gravitational collapse
at the end of its life cycle on exhausting its internal
nuclear fuel. It would be then of much physical interest
to investigate the astrophysical consequences of their formation,
also taking into account the possible quantum gravity effects
these may cause
\cite{Goswami},
and the possible connection of the same with very highly
energetic astrophysical phenomena, such as the gamma ray bursts
and those related to the active galactic nuclei.

We show here that the naked singularities forming in
gravitational collapse could provide us with a window into
the new Planck scale physics, even far away from the
actual singularity. This is because, the particles
could go very close to a naked singularity and then emerge
with very high velocities near the Cauchy horizon
that the naked singularity created.

In the next Section II, we discuss the naked singularity and
black hole formation in a general spacetime universe, and clarify the
basic difference between these two fundamentally different final
outcomes of a complete gravitational collapse of massive matter
clouds in general relativity. Section III then investigates
and shows the divergence of center of mass energy in particle
collisions near the Cauchy horizon, between the very high-energy
particles coming from vicinity of the naked singularity, and the other
ingoing particles. In Section IV, we examine and consider different
physical possibilities and mechanisms as to how a naked singularity
would possibly generate ultra-high energy particles, emerging
from its vicinity.
In Section V  we then construct a toy collapse model to
describe
and demonstrate the ideas presented in Section IV. We show that
when the repulsive nature of gravity in the vicinity of naked
singularity is modeled by a large negative pressure,
we naturally get highly energetic outgoing particles
in the vicinity of Cauchy horizon that participate in high
energy collisions.
The final Section V gives a few concluding remarks
and the possible future line of investigations.

\section{Gravitational collapse to  naked singularity}

We now consider a general gravitational collapse scenario,
where the collapse of a matter cloud proceeds from regular
initial data, specified on an initial spacelike hypersurface.
As a result of the complete gravitational collapse, a spacetime
singularity is formed, and depending on the nature of the
matter initial data and the velocities of the collapsing
shells, the singularity could be either covered within a
black hole, or it could be a naked singularity, where the
event horizon does not hide the curvature divergence.

In the latter case above, where the evolution of the
data in the complete gravitational collapse as described by Einstein equations
results in the formation of naked singularity, the presence
of the naked singularity is characterized
by the existence of families of future directed outgoing null and
timelike geodesics, which terminate in the past at the singularity,
and in future they reach a faraway observer in the spacetime.

When the gravitational collapse ends in a naked singularity,
this does not mean that the causality of the spacetime must be
necessarily broken. The collapse may develop into a naked
singularity but there need not be any closed timelike curves
in the spacetime, and in that sense the spacetime is fully regular.
This can be seen clearly in the dust collapse models, and many
other gravitational collapse scenarios, where a
naked singularity of collapse develops, and which have been
analyzed in detail over past years
\cite{dust}.

In such a case, the spacetime necessarily admits a {\it Cauchy
horizon}, which can be defined by the first null ray coming out
of the singularity, in the case of spherical symmetry. In general,
the future light cone of the very first point on the singularity
curve is the Cauchy horizon in the spacetime (see Fig.1).
\begin{figure}
\begin{center}
\includegraphics[width=0.5\textwidth]{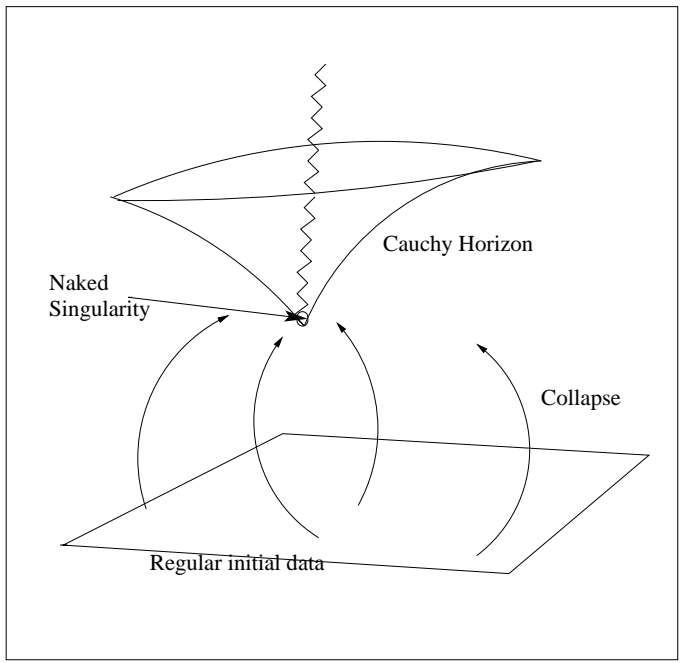}
\caption{\label{fig1}
The continual gravitational collapse starting from a
regular initial data on a partial Cauchy hypersurface results in
the formation of a naked singularity. The boundary of future light
cone of the first point of naked singularity is the Cauchy
horizon.}
\end{center}
\end{figure}

When the collapse terminates in a naked singularity, the
spacetime is no longer {\it globally hyperbolic}, in the sense
that there does not exist any global spacelike surface, such that
the initial data on the same would fully predict and determine
the future as well as past evolutions in the spacetime for
all times (for more details on global hyperbolicity and
causal structure properties, see e.g. \cite{HE}).

In that case, it is the Cauchy horizon which marks the
boundary of the region of the spacetime which is fully predictable
from an initial surface. A naked singularity is therefore always
characterized necessarily by the presence of a Cauchy horizon,
which is in general a three dimensional null hypersurface in
the spacetime, generated by null generators.

\section{Divergence of center of mass energy}

We now consider the collisions between particles, the
outgoing ones being those that emanate from close to the
ultra-strong gravity region, that is the naked singularity,
with those which are the infalling particles. The underlying
scenario is that of a gravitational collapse that results
in a naked singularity, and the collisions take place in
the vicinity of the Cauchy horizon that is created by
the naked singularity, and which is the first light cone
coming out of the naked singularity. Our methodology
is similar to that used in
\cite{Zasla3},
to study the particle acceleration near an
event horizon, which is also a three-dimensional null
hypersurface in the spacetime.

Let $l_{\alpha}$ be a null vector field on the
spacetime,
which we take to be a generator of the Cauchy horizon of
the spacetime that admits a naked singularity.
Thus we have,
\be
l_{\alpha}l^{\alpha}=0.
\label{eqn1}
\ee
We choose another null vector field $n_{\alpha}$, thus satisfying
\be
n_{\alpha}n^{\alpha}=0.
\label{eqn2}
\ee
We normalize $n_{\alpha}$ with respect to $l_{\alpha}$
so that
\be
n_{\alpha}l^{\alpha}=-1.
\label{eqn3}
\ee.
We can then choose two mutually orthogonal spacelike vector fields
$b_{i}$, $i=1,2$
\be
b_{i \mu}b_{j}^{\mu}=\delta_{ij},
\ee
which are orthogonal to the null vectors $l_{\alpha},n_{\alpha}$.
Thus,
\be
l_{\alpha}b_{i}^{\alpha}=0 ,n_{\alpha}b_{i}^{\alpha}=0 .
\ee

In terms of the four vector fields
$l_{\alpha}$,$n_{\alpha}$,$b_{1\alpha}$,$b_{2\alpha}$, the spacetime
metric $g_{\mu\nu}$ can then be written as,
\be
g_{\mu\nu}=-l_{\mu}n_{\nu}-n_{\mu}l_{\nu}+\sigma_{\mu\nu},
\label{eqn4}
\ee
where
\[ \sigma_{\alpha \beta}= b_{1\alpha}b_{1\beta}+ b_{2\alpha}b_{2\beta} \]
is a spatial metric on the two-surface orthogonal to null
vectors $l_{\alpha},n_{\alpha}$
(see \cite{Zasla3} and \cite{Poisson1}).
Note that
\be
\sigma_{\alpha \beta}l^{\alpha}=0,\sigma_{\alpha \beta}n^{\alpha}=0.
\ee

The velocity $U^{\mu}$ of a particle at any point in
the spacetime can be written down now as a linear combination of
these four vectors $l,n,b_{1},b_{2}$, since they form a basis
for the tangent vector space,
\be
U^{\mu}=A \left(\frac{l^{\mu}}{\alpha}+ \beta n^{\mu}+ \gamma^{\mu}\right),
\label{eqn5}
\ee
where
\[\gamma^{\mu}=c_{1}b_{1}^{\mu}+c_{2}b_{2}^{\mu}\]
and $A$ is a normalization factor chosen so that the
velocity vector is normalized $U^{\mu}U_{\mu}=-1$.
From \eq{eqn1},\eq{eqn2},\eq{eqn3},\eq{eqn4},\eq{eqn5},
we can write,
\be
A=\frac{1}{\sqrt{\frac{2\beta}{\alpha}-\sigma_{\alpha \beta}
\gamma^{\alpha}\gamma^{\beta}}}.
\ee

We now consider two particles of masses $m_{1},m_{2}$,
traveling with velocities $U_{i}$,$i=1,2$, which are given by,
\be
U_{i}^{\mu}=A_{i} \left(\frac{l^{\mu}}{\alpha_{i}}+ \beta_{i} n^{\mu}
+ \gamma_{i}^{\mu}\right),
\label{eqn6}
\ee
with
\[
A_{i}=\frac{1}{\sqrt{\frac{2\beta_{i}}{\alpha_{i}}
-\sigma_{\alpha \beta}\gamma_{i}^{\alpha}\gamma_{i}^{\beta}}}.
\]
When these particles interact at a given spacetime point,
the energy of collision in the center of mass frame is given by
(see \cite{Patil} and \cite{BSW}),
\be
E_{cm}^2= m_{1}^2+m_{2}^2- 2m_{1}m_{2}g_{\mu\nu}U_{1}^{\mu}U_{2}^{\nu}.
\label{eqn7}
\ee
From \eq{eqn1},\eq{eqn2},\eq{eqn3},\eq{eqn4},\eq{eqn6},\eq{eqn7},
we then finally obtain,
\be
E_{cm}^2= m_{1}^2+m_{2}^2+ \break
\frac{2m_{1}m_{2}\left( \frac{\beta_{1}}{\alpha_{1}}+\frac{\beta_{2}}{\alpha_{2}}
-\sigma_{\alpha \beta}\gamma_{1}^{\alpha}\gamma_{2}^{\beta}\right)}
{\sqrt{\frac{2\beta_{1}}{\alpha_{1}}-\sigma_{\alpha \beta}
\gamma_{1}^{\alpha}\gamma_{1}^{\beta}}\sqrt{\frac{2\beta_{2}}{\alpha_{2}}
-\sigma_{\alpha \beta}\gamma_{2}^{\alpha}\gamma_{2}^{\beta}}}.
\label{eqn8}
\ee

\begin{figure}
\begin{center}
\includegraphics[width=0.5\textwidth]{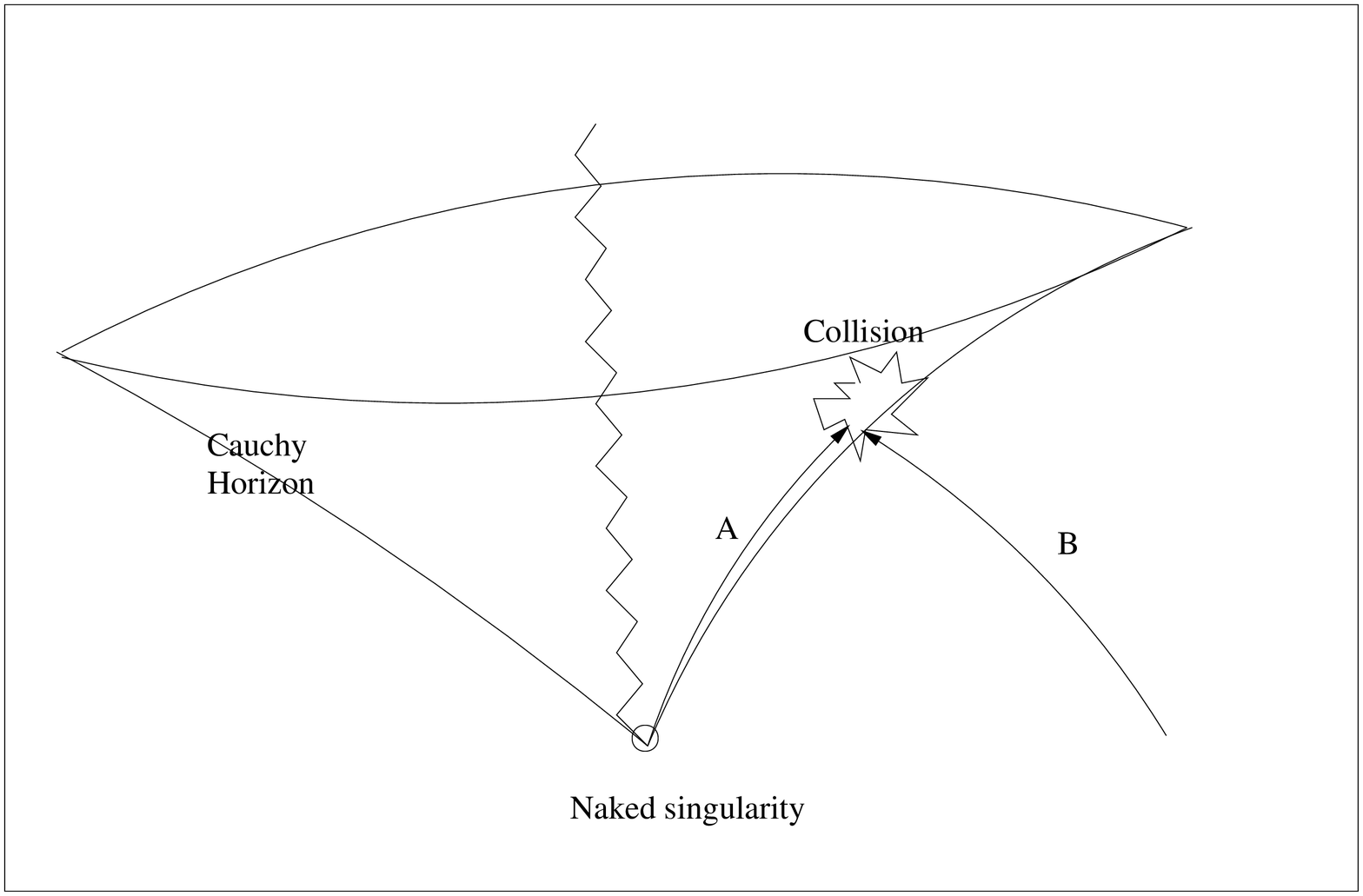}
\caption{\label{fig2}
A highly energetic particle "A" that has been accelerated
by the naked singularity travels close to the Cauchy horizon.
When it collides with an incoming particle "B", the center of mass
energy is arbitrarily large, depending on how close is the point
of collision to the Cauchy horizon. }
\end{center}
\end{figure}

We have considered here a situation where two particles
collide near the Cauchy horizon(see Fig.2).
One of the particles is an ingoing particle, and the other one
is a highly energetic outgoing particle, which is coming
from a close neighborhood of the singularity, and which therefore
travels close to the Cauchy horizon. Such a particle
can arise in a following way, as we described in our previous paper
\cite{Patil}.
One can consider the region of spacetime
before the formation of singularity, and the ingoing geodesics
starting from a faraway region. After passing through the
regular center these would emerge as outgoing geodesics. The outgoing
particle close to the Cauchy horizon would be the ingoing geodesic,
which just missed the naked singularity and emerged as an
outgoing particle. Such ingoing particles in collapse may
miss the singularity, if they had a small angular momentum
or due to the small perturbations in geometry.

This particle (say particle 1) stays very close
to the Cauchy horizon, and thus travels almost in the
direction of the horizon generator $l_{\mu}$.
For this to happen we must have
\be
\alpha_{1} \rightarrow 0.
\label{eqn9}
\ee
Thus it can be clearly seen that the center of
mass energy \eq{eqn8} diverges in the limit \eq{eqn9}
\be
E_{cm}^2 \sim \frac{1}{\sqrt{\alpha_{1}}} \rightarrow \infty.
\ee
The value of $\alpha_{1}$ signifies how close is an
outgoing particle to the Cauchy horizon at the point of
collision. Smaller the value of $\alpha$, larger is the
center of mass energy of collision between the ingoing and
outgoing particles. Thus, by making $\alpha$
arbitrarily small, center of mass energy of collision
can be made arbitrarily large.

At this point we would like to compare the results
here with that of the previous work \cite{Patil}.
Earlier, we showed the divergence of center of mass energy
between two identical massive particles following timelike
geodesics, one of them ingoing and the other one outgoing.
We assumed that the spacetime admitting naked singularity was
spherical symmetric and also self-similar. These additional
constraints imposed on the spacetime allowed us to integrate
the geodesic equations and to explicitly write down the
expression for the velocity of the particles. Then one
can compute the center of mass energy, which was divergent
in the limit where the point of collision approached
the Cauchy horizon.

In the treatment here, we have considered the collision
between two particles where their masses need not be the same.
The condition that the particles follow geodesic motion
is not imposed. This allows us to go beyond the test particle
approximations, and consider a collision between particles
that constitute the fluid which is a source of curvature of
spacetime. As will be discussed later, the highly repulsive
nature of the naked singularity, when quantum gravity effects
are accounted for, could eject the initially collapsing
fluid elements in radially outward direction with extraordinarily
high energies. Particles constituting the fluid element would
then travel close to what would have been the Cauchy horizon
at classical level. These would then collide with incoming
particles, which may be either the test or fluid particles.
Since the fluid elements do not follow the geodesic motion in
the presence of pressures, going beyond the assumption that
the colliding particles follow geodesic motion would be useful.
In fact, the only requirement necessary for the divergence
of center of mass energy now is that, the velocity of one of
the colliding  particles, when expanded in the basis
$\left(l^{\mu},n^{\mu},b_1^{\mu},b_2^{\mu}\right)$, gets a dominant
contribution from $l^{\mu}$, which approaches the generator
of the Cauchy horizon in the limit of the particle being
closer and closer to the Cauchy horizon.
This is again necessarily the requirement of the existence
of particles with higher and higher energies, which would then
travel closer and closer to the Cauchy horizon, coming
from the vicinity of the naked singularity.

\section{Physical mechanisms of generation of energetic particles}

Collisions with arbitrarily high center of mass
energies require the existence of extremely energetic
particles, traveling close to the Cauchy horizon, which
come from the vicinity of a naked singularity. In this section, we
would like to discuss various mechanisms and physical
scenarios in which such particles would be generated.
The possible repulsive nature of gravity, as we discuss
below, can play a crucial role in such
acceleration mechanisms.

In general, gravity is always thought to be
attractive. This fact is based on the purely Newtonian intuition
rooted in our daily experience. However, gravity
can exhibit a repulsive nature as well. Gravity turns out
to be repulsive when the stress-energy tensor of the matter
field, which is the source of spacetime curvature, violates
the weak energy condition. The weak energy condition is violated
when the pressures are sufficiently negative as compared to the
energy density of the matter. In general relativity, since
not only energy density but also the pressures gravitate,
large negative pressures give rise to the repulsive gravity.
The well-known examples of this phenomenon being the
accelerated expansion of the universe due to a cosmological
constant, and the inflation in early universe.

Similarly, the gravitational singularities that
occur in general relativity, that are characterized by
the divergence of Kretschmann scalar invariant
$I=R_{\alpha \beta \gamma \delta} R^{\alpha \beta \gamma \delta}$,
are commonly thought to be intrinsically and
infinitely attractive in nature. However, the
spacetime singularities can also exhibit a repulsive
nature at times. We argue below that, as for the particle
acceleration from the vicinity of singularities, an
important role could be played here by the repulsive
nature of naked singularities, both at the classical and
quantum levels, in the process of acceleration of such
particles. This, however, need not be the only mechanism
to generate very high energy particles emanating from
the vicinity of naked singularities.

\subsection{Classical considerations}

As we have already discussed earlier,
infalling particles from infinity which just missed
the naked singularity would emerge as highly energetic
ultra-relativistic outgoing particles, and would travel
close to Cauchy horizon. If the naked singularity that
formed shows a repulsive behavior at classical level,
particles then get further boosted up in energy,
which we discuss in some detail in this section.

Gravity can show a repulsive character in the vicinity
of naked singularities, although the weak energy conditions
might still be respected. Such a repulsive nature, at the
level of classical general relativity, is one of the
intriguing and remarkable features associated with naked
singularities. There are several ways to define and characterize
the repulsive nature of naked singularities. Many of these are
specific to the coordinate system chosen. However, recently
there has been a significant progress towards the coordinate
invariant definition of repulsive nature of gravity
(see \cite{Felice}, \cite{Felice1}, \cite{Luongo}).

One can use the motion of test particles to study the
repulsive nature of naked singularities. In stationary axially
symmetric spacetimes, the radial motion can be described in terms
of the effective potential, the explicit form of which depends on
the coordinate system chosen. The effective potential has been
demonstrated to be repulsive in the vicinity of naked singularities
in the case of Reissner-Nordstr\"{o}m, Kerr, and Kerr-Newmann
naked singularities, which are inside the inner horizon of
black-holes. This is also the case for naked
singularities in these spacetimes, in the superextremal
cases where the horizon is absent, thus exposing the
singularity to the observer at infinity.
It has also been suggested that the
widening, instead of shrinking of the light-cones in approach to
the naked singularity in the coordinates system that is adapted
to the spacetime symmetries would be the indicator of repulsive
nature of gravity \cite{Felice1}. This conjecture has been verified
in the cases mentioned above as well. While the two approaches
to characterize repulsive nature of gravity around naked singularities
were observer dependent, an invariant way to define a repulsive
gravity was proposed recently \cite{Luongo}. This involves
the analysis of the eigenvalues of the curvature in the
SO(3,C) representation. The oscillation of the eigenvalues,
while approaching the naked singularity along an arbitrary
spatial direction marks the onset of the repulsive nature
of gravity. This criterion has also been demonstrated to be
consistent with cases discussed above in the context of other
approaches. Thus gravity has been shown to exhibit repulsive
nature in the vicinity of naked singularities in the stationary
axially symmetric spacetimes .

While there has been an extensive analysis of repulsive
nature of naked singularity in stationary spacetimes, there is
not much attention paid to the singularities formed in the
gravitational collapse. This is perhaps because so far the
analysis has been dependent on the reference frames adapted
to the spacetime symmetries, specifically the time translation
symmetry, to analyze the geodesic motion of test particles.
In the absence of a timelike killing vector in the case of
gravitational collapse models, such an analysis is nontrivial
and would be plagued by the gauge related issues. However, with
the advent of gauge invariant techniques it might be possible
to analyze the repulsive nature of gravity in non-stationary
naked singularity models. We are currently investigating
this issue, and would like to present our findings in
future elsewhere.

At this point, however, we would like to conjecture
that the naked singularities formed in the realistic gravitational
collapse models would exhibit a repulsive nature in their vicinity,
in the framework of classical general relativity. This would then
play an important role in the particle acceleration mechanism
in their vicinity. A test particle infalling from infinity,
marginally missing the naked singularity, and thus emerging as
a highly energetic outgoing particle, would slow down very soon
as it loses energy. Then it would deviate from its motion close
to the Cauchy horizon, if gravity is not repulsive in the
vicinity of the singularity. Thus, the desired collisions with
arbitrarily large center of mass energies would occur in the
small neighborhood around singularity. However, if gravity is
repulsive then the outgoing particle would be further accelerated,
and can thus continue to be highly energetic even in the
region faraway from the singularity, as it travels closely
near to the Cauchy horizon direction. Thus the collisions with
arbitrarily high center of mass energies would happen in the
region sufficiently faraway from the naked singularity, but
close to the Cauchy horizon.

In a gravitational collapse starting from regular initial
data resulting in the naked singularity, one might expect the
transition where gravity would change its nature from being
attractive to repulsive. In the beginning in the absence of
singularity the gravity would be attractive. But on the onset
of formation of naked singularity it would turn repulsive. It
would be very interesting to study such a change in nature.
It would play influential role in accelerating the test particle,
when it is infalling, as well as when it travels outwards barely
missing the singularity, by providing a further kick to
provide a further acceleration to the particle so that it
continues to travel faraway with high energy close to
the Cauchy horizon.

\subsection{Quantum gravity effects near naked singularity}

Having discussed the role repulsive gravity around naked
singularities might play in the mechanics of particle acceleration,
we now consider the repulsive aspects associated with quantum
gravity. Classical general relativity provides us with a good description
of gravitational collapse up to the stage when densities and
curvatures are small as compared to the Planck scale.
But at Planck scales, the classical dynamics is no more a
valid description of gravitational collapse, because the
quantum mechanical effects must be taken into consideration
as well. Thus we need a consistent quantum theory of gravity
to deal with this situation. The spacetime is necessarily
discrete or with a complicated topology at a deep down
fundamental level. However, around the Planck scale in
a certain regime, effective dynamics of the collapse can
be at times described by modifying the Einstein equations,
incorporating the non-perturbative corrections due to
quantum gravity.

\begin{figure}
\begin{center}
\includegraphics[width=0.5\textwidth]{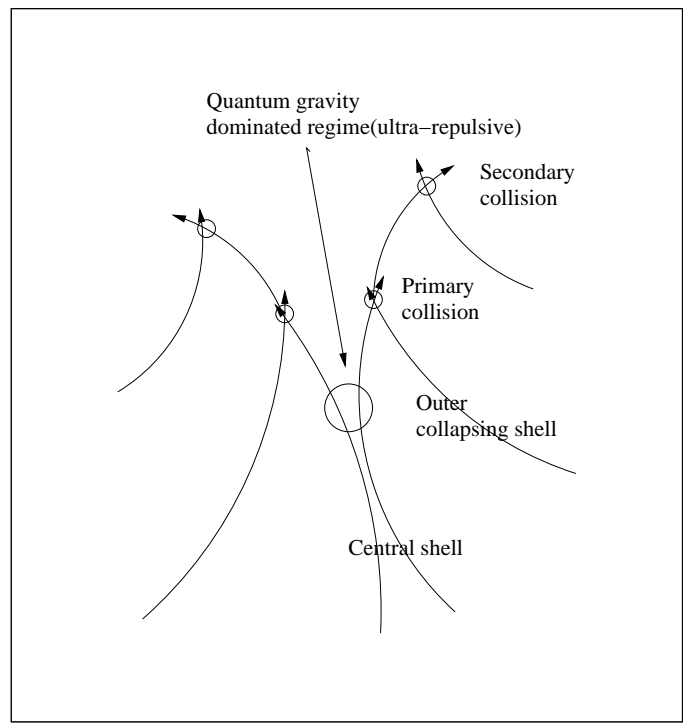}
\caption{\label{fig3}
A quantum gravitational collapse with resolution of
the classical naked singularity. The central shell undergoes a
collapse and rebounce, having gone through the repulsive
quantum gravity dominated regime. When highly energetic particles
traveling outwards along the classical Cauchy horizon collide
with the particles from outer collapsing shells, there are
primary and then secondary collisions at arbitrarily high
center of mass energies. A fireball is created, of finite
extent, which is dominated by the Planck scale physics.
}
\end{center}
\end{figure}

There have been a few attempts is this direction.
We shall briefly discuss here one such work, describing the
evolution of the collapsing object which would have classically
led to a naked singularity, but taking into account the
corrections due to quantum gravity resolves and removes
the naked singularity
\cite{Goswami}.
In this work, the quantum theory of gravity that
is used to modify classical dynamics is the Loop quantum gravity
formalism. This implements the non-perturbative quantization of
gravity rewritten in Ashtekar variables in a consistent way.
Loop quantum gravity has already successfully dealt with and
resolved issues like singularity avoidance in cosmology
\cite{Bojowald},
and such others. The model which gives rise to a naked
singularity at classical level consists of a homogeneous scalar
field with a suitable potential, with a finite spatial extent,
which is matched to a generalized Vaidya exterior region
at a constant comoving radius in a comoving reference frame.
The potential is chosen in such a way that during the collapse
the pressure is sufficiently negative so that the gravitational
mass at any point in the spacetime is smaller than that of
the physical radius in natural units. Therefore the trapped surfaces
are not formed as the collapse evolves, and also there is an
outgoing flux of energy in the Vaidya region. As the collapse
proceeds, the singularity is formed in a finite proper time,
as the density and curvatures blow up to infinity. The singularity
is naked due to the avoidance of the trapped surfaces formation
and there is an outward energy flux from the singularity
which can be seen from infinity.

This classical scenario is modified by taking into
account the non-perturbative semi-classical modification based
on Loop quantum gravity. The dynamics at the semi-classical level
which is valid in a certain regime around the Planck scale
consists of modification of the matter Hamiltonian, thus
modifying the spacetime density and pressures in the Einstein
equations. The effective pressure is highly negative in comparison
to the effective energy density and the effective equation of
state for matter violates the weak energy conditions by an
enormously large margin. The effective equation of state on
the other extreme of validity of semi-classical regime turns out
to be $\rho_{eff}=-9 p_{eff}$, whereas the weak energy condition
demands $\rho_{eff}+p_{eff} \geq 0$.
This result is shown to be true for other
matter fields as well
\cite{Singh}.
There is a large outburst
of energy, in the very final stages of the collapse,
due to the extremely large negative pressure. As was
mentioned earlier, the violation of energy conditions implies
the repulsive nature of gravity. Since the pressure in this
case is highly negative, gravity is ultra-repulsive. The
density goes on decreasing, instead of blowing up like in the
classical case, and the scale factor remains non-zero and
well above the classical value till the breakdown of semiclassical
regime. Beyond this regime the evolution would be governed
by quantum difference equations.

Although the scenario described above is a toy model,
describing the modified collapse dynamics of a homogeneous scalar
field cloud with exterior Vaidya region, it gives an important
insight so as to what the final collapse outcome might be
in a generic collapse scenario following the complete quantum
dynamics beyond the semiclassical regime. In general, it would be
reasonable to expect that the naked singularity of collapse
would be resolved due to quantum gravity effects. Due to
ultra-repulsive nature of gravity in semi-classical and quantum
regime, the fluid elements in the innermost core of the cloud
that would have otherwise collapsed to the naked singularity
for the first time, would be now thrown out and there would
be large outburst of energy in the outward direction along
what would have been classically the Cauchy horizon.

The particles constituting this ejected central fluid
element would be highly energetic and travel at ultra-relativistic
speeds along the Cauchy horizon. These, then would collide
with the incoming particles, which could be either test particles
or constituents of outer collapsing fluid elements, and the center of
mass energy of such collisions would then be arbitrarily large
(see Fig.3).
The primary collisions would produce secondary relatively
less energetic particles, which would be colliding with other
particles with sufficiently high center of mass energies.
Thus the collisions with divergent center of mass energies along
the Cauchy horizon would generate a fireball of finite extent
which would be dominated by quantum gravity effects
and physics at ultra-high energies below the Planck scale,
but still well above the center of mass energies realized in
the terrestrial particle accelerators.

\begin{figure}
\begin{center}
\includegraphics[width=0.5\textwidth]{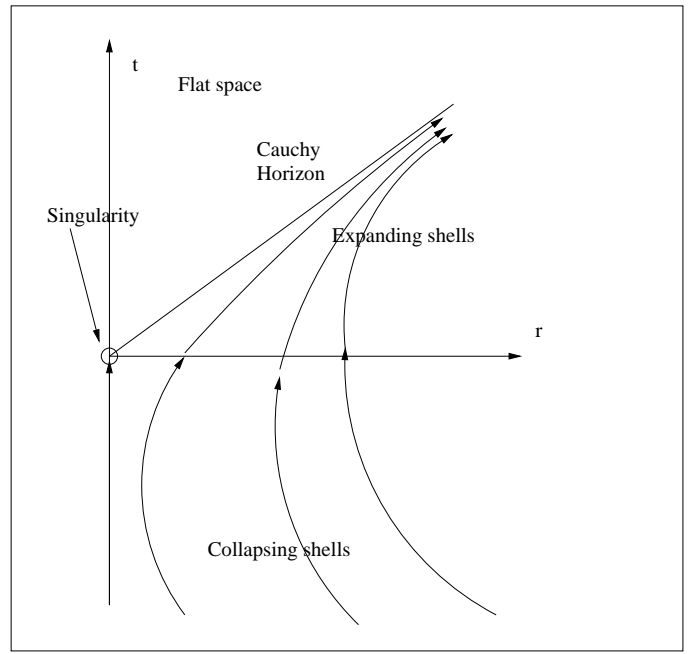}
\caption{\label{fig4}
Gravitational collapse of a fluid with large pressure
inhomogeneities eventually turns into a dispersal with fluid
elements moving outwards, eventually reaching closer and closer to
the speed of light. The central shell, however, collapses to
form a momentary singularity, which eventually disappears.
All the non-central fluid elements asymptotically approach the
Cauchy horizon of a naked singularity. When an incoming test
particle interacts with a highly energetic fluid particle,
the center of mass energy of collision is arbitrarily large.
This is a schematic diagram describing this situation in
the Schwarzschild-like coordinates.}
\end{center}
\end{figure}

\subsection{Outward acceleration due to inhomogeneities}

There is an another interesting possibility
where one might get highly energetic collisions along
the Cauchy horizon (see Fig.4).
Consider gravitational collapse of a fluid with non-vanishing
pressure, for example, with a linear equation of state $p=k\rho$
with a sufficiently large $k$. Also consider a case of inhomogeneous
collapse, where the density and pressure decrease as we move outwards
from the center of collapse. There would then be a gradient of
pressure pointing in the radially outward direction. Thus there
is an outward force on the fluid element, thus accelerating
it outwards. The density goes on increasing during the gravitational
collapse, and if the distribution of matter is sufficiently
inhomogeneous, then the pressure gradient also builds up.
The fluid elements then experience larger and larger outward
force. The central shell, however, undergoes a continual
collapse since the regularity conditions demand the vanishing
of pressure gradient at the center. Thus the center eventually
collapses to singularity. However, if the trapped surfaces are
not formed, then the non-central fluid shells eventually
bounce back even before their density reaches the Planck scale
since the density and pressure gradients build up and attain
large values and the large outward force causes the fluid
elements to expand, following a contracting phase.
All non-central shells accelerate outwards, reaching
arbitrarily high velocities, and thus approach the Cauchy
horizon of the central naked singularity. Solutions such as
these are described, for example, in \cite{Ori} within the
context of self-similar models, and have been referred
to as explosive solutions.
In cosmology, such an effect has also been used
to explain the observed cosmic acceleration
of the universe, in the context of inhomogeneous
scalar field models
\cite{Nakao}.

Infalling test particles from infinity which
just miss the singularity are accelerated till they cross
the momentary singularity in this case, and thereafter
they would emerge as highly energetic outgoing particles
in the Minkowski patch. They would then retain their energy,
and thus continue to be ultra-relativistic in the
region close to Cauchy horizon, till infinity.

When an incoming test particle interacts with
the highly energetic fluid particle, or the outgoing
test particle, in the vicinity
of the Cauchy horizon, then in that case the center
of mass energy of the collision is arbitrarily
large, depending on how close is the point of
interaction to the Cauchy horizon.

\section{Mass-Energy emission from the vicinity of a timelike naked singularity}

In the previous section we described various mechanisms of
generation of highly energetic outgoing particles in the vicinity
of a naked singularity that travel close to the Cauchy horizon
and participate in the high energy collisions. We argued that the
the very nature of the naked singularity,
at either the classical or semiclassical level,
where large negative repulsive pressures could arise due to quantum
gravitational corrections to the Einstein`s equations,
would generate highly energetic particles.

In this section we construct a gravitational collapse
model to demonstrate these ideas.
The collapse here describes the
dynamical evolution
of a matter cloud in the final stages of collapse
leading to a timelike naked singularity.
The cloud consists of a perfect fluid with a variable
equation of state which is dependent both on time as well
as spatial coordinates. Such a scenario appears to be
reasonable in view of the fact that the collapse being
a dynamical process leading to very high densities,
the equation of state describing the relation between
pressure and density in the collapsing cloud, which
would be essentially dictated by a physical theory valid
at that energy scale, is not to be expected to remain
unchanged as collapse progresses.

Our model given below is purely classical and we assume
that the pressure turns negative during the final stages
of collapse, close to the formation of singularity. Negative
pressure can either be thought
of as a model to describe an onset of a classical
repulsive regime near the singularity, or it could also
be interpreted as an indication
of an occurrence of quantum gravity effects. We also
assume that the spacetime is spherically symmetric since
it would allow us to shed a light on the role played by
a repulsive naked singularity in the particle
acceleration process in a rather transparent way.

We shall see that under some circumstances the mass
of the core of the collapsing star is radiated away by means
of the negative pressures that originate near the center
as the density approaches the singularity.
Since the spacetime is spherically symmetric, no energy
can be carried away in the form of gravitational waves. Therefore,
if a finite amount of the infalling mass is ejected from
the collapsing cloud, it must be in the
form of photons or massive particles. In the final
stages the cloud will therefore be matched to a generalized
Vaidya exterior geometry, describing a space-time filled
with outgoing radiation
\cite{matching}.
The outgoing radiation represents the energy
carried out from the evaporation process of the
central mass.
At the time of formation of the singularity the entire
mass of the cloud is radiated away, and therefore after the
formation of the singularity the exterior spacetime
will be described by the Minkowski metric.
Particles escaping the vicinity of the singularity
by this mechanism would have to be intrinsically
highly energetic, since they originate
in the region with extremely large curvature where
quantum gravity is dominant and they would travel
outwards in the vicinity of the Cauchy horizon
(See Fig.5).

\begin{figure}
\begin{center}
\includegraphics[width=0.5\textwidth]{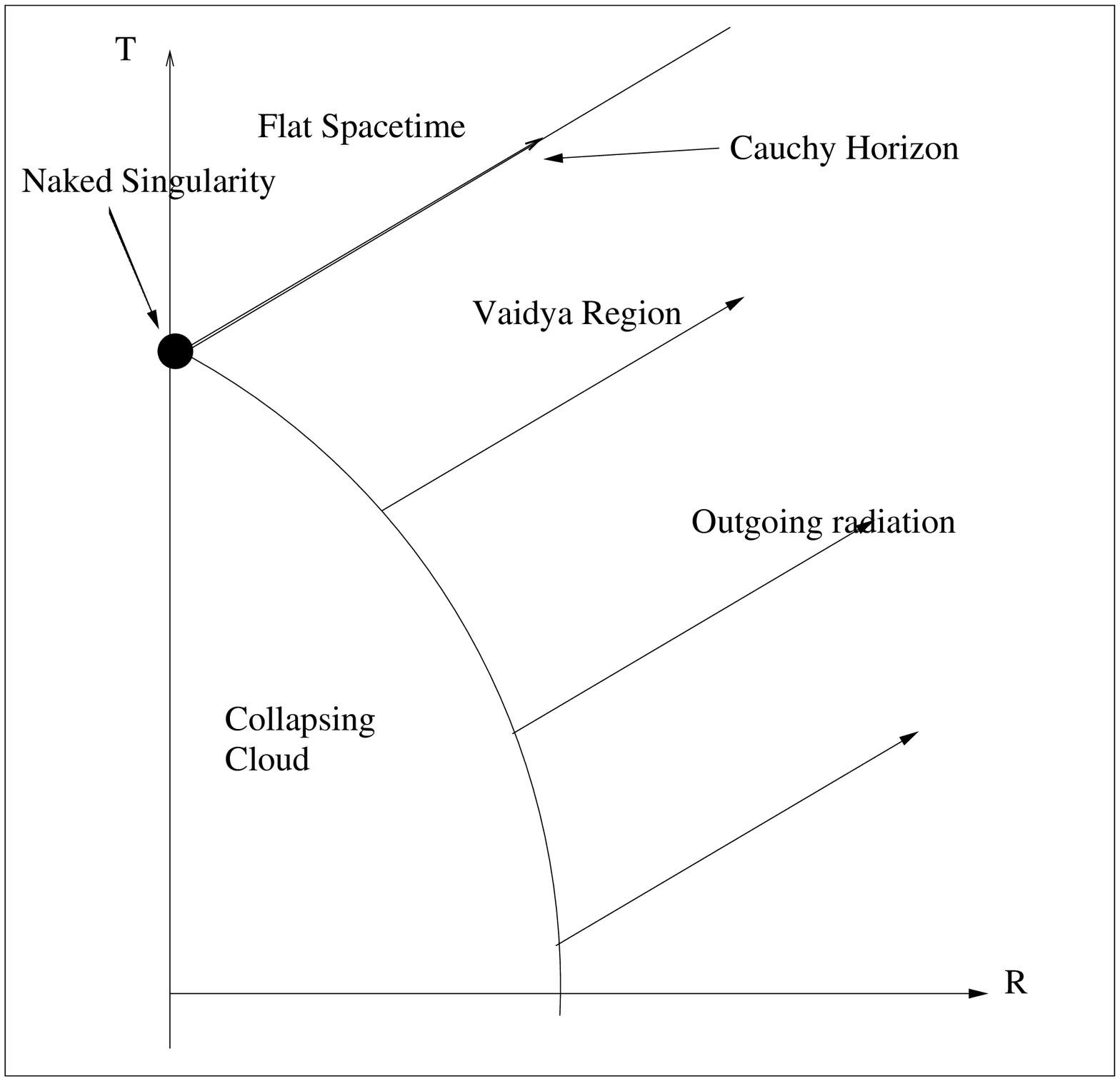}
\caption{\label{fig5}
A schematic diagram of gravitational collapse
of a matter cloud to a naked singularity in Schwarzschild
like coordinates $(T, R)$. The mass is radiated away
due to the negative pressure in the late stages of collapse.
The region outside the matter cloud is described by
a generalized Vaidya metric. The cloud is radiated
away completely leaving behind a flat spacetime.
The same picture, given in comoving coordinates $(t,r)$,
would depict the naked singularity as an extended
timelike curve.
}
\end{center}
\end{figure}

We will assume that
\begin{equation}\label{eos}
    \frac{p(r,t)}{\rho(r,t)}=k(r,t)
\end{equation}
where $r$ and $t$ are the comoving radius
and time of the collapsing matter shells.
In the following, primed quantities will denote
derivatives with respect to $r$ and dotted quantities
will denote derivatives with respect to $t$.

From the Einstein equations we see that
\begin{eqnarray}\label{p}
  p &=& -\frac{\dot{F}}{R^2\dot{R}} \\ \label{rho}
  \rho &=& \frac{F'}{R^2R'}
\end{eqnarray}
where $F$ is the Misner-Sharp mass of the system
which is related to the amount of energy enclosed in the
comoving radius $r$ at the time $t$, and $R$ is the physical
radius of the shell that is shrinking to zero size.
The collapse scenario is described by the condition
that $\dot{R}<0$.

The Misner-Sharp mass of the system is then defined by
\begin{equation}\label{F}
    F=R(1-G+H)
\end{equation}
where $G$ and $H$ are the metric functions.

The spacetime metric in terms of functions $G(t,r),H(t,r)$ and
$R(t,r)$can be written as
\begin{equation}
ds^2=-\frac{\dot{R}^2}{H}dt^2+\frac{R^{\prime 2}}{G}dr^2+R^2\left(d\theta^2+sin^2 \theta d\phi^2\right)
\end{equation}

The remaining Einstein equations can then be written as
\begin{eqnarray}\label{G}
  2\dot{R}' &=& R'\frac{\dot{G}}{G}+\dot{R}\frac{H'}{H}  \\ \label{H}
  \frac{p}{\rho+p} &=& -\frac{1}{2}\frac{R'}{\dot{R}}\frac{\dot{G}}{G}
\end{eqnarray}

We therefore have a system of six equations
(the equation of state, the definition of the Misner-Sharp mass
and the four Einstein equations), in seven unknowns
($p, \rho, k, R, F, G, H$), therefore resulting in the
freedom to choose one function
\cite{trapped}.
Once $k$ is chosen, we
can obtain $F$, $p$ and $\rho$ as functions of
$R$ from equations \eqref{eos}, \eqref{p} and \eqref{rho}.
Then equations \eqref{G} and \eqref{H} can be used
to obtain $G$ and $H$. The whole system then reduces to
solving the equation of motion given by \eqref{F},
which can be written as
\begin{equation}\label{mot}
    \dot{R}=\pm\frac{k+1}{k}\frac{R'\dot{G}}{HG}\left(\frac{F}{R}+G-1\right)
\end{equation}

Before we proceed we shall note two points.
First of all the whole scenario has still a scaling
degree of freedom, therefore by defining a function $v(r,t)$
from $R=rv$ and fixing the scale at the initial time
to be $R(r,t_i)=r$, we can substitute the function $R$
with $v$. Collapse will then be obtained for $\dot{v}<0$,
and the singularity is achieved for $v=0$. At this point we note
that approaching the singularity, $v$ is monotonically
decreasing in time, therefore allowing us to invert $v(r,t)$
and consider it as an alternative time coordinate.
This represents the time at which the shell labeled by $r$
reaches the event $v$. We shall therefore consider all
functions as depending on $r$ and $v$ and rewrite the equations
accordingly. The equation of motion, expressed in
term of $\dot{v}$, can then be inverted as well
to give $t(r,v)$.

Regularity of the matter model at the center of
the cloud imposes some restrictions on the possible
behaviors of the Misner-Sharp mass. It can be easily
checked that we must have
\begin{equation}
    F=r^3M(r,v)
\end{equation}
and that $M$ may not contain linear terms
in $r$ in order to avoid cusps of the density
at the center.

From the equation of state and equations
\eqref{p} and \eqref{rho}, we derive the differential equation
that must be satisfied by $M$ as,
\begin{equation}\label{diff}
    3kM+krM_{,r}+\left[v+(k+1)rw\right]M_{,v}=0
\end{equation}
where $M_{,r}$ denotes derivatives of $M$ with
respect to $r$ in the $(r,v)$ coordinates,
and $w(r,v)$ is given by $v'$ expressed as a function
of $r$ and $v$ ($w=v'(r,t(r,v))$).
We shall note here that in principle $w$ is
not known because it requires the knowledge of $v$
which comes from the integration of the equation of
motion \eqref{mot}. Nevertheless in certain cases,
corresponding to some specific
choice of $k$, it might be possible to evaluate $M$.

We choose the function appearing in the equation
of state, which is $k(r,v)$, as the free function of
the system. Since all the quantities involved are well
behaved (and at least $\mathcal{C}^2$) outside the
singularity, we can
always perform an expansion near the center and write
\begin{equation}
    k(r,v)=k_0(v)+k_1(v)r+k_2(v)r^2+...
\end{equation}
This implies that $p,$ $\rho$ and $M$ can be expanded accordingly.
Restricting ourselves in a close neighborhood of the
center (as it is relevant for our purposes) we can expand
equation \eqref{diff} and rewrite it, equating term by term.
We thus obtain the following set of differential equations,
\begin{eqnarray}
  0 &=& 3k_{0}(v)M_{0}(v)+M_{0,v} v  \\
  0 &=& 3k_1M_0+4k_0M_1+M_{1,v}v+(1+k_0)w(0,v)M_{0,v}  \\ \nonumber
  0 &=& 3k_2M_0 4k_1M_1+5k_0M_2+M_{2,v}v+k_1M_{0,v}w(0,v)+\\
  &+& (1+k_0)M_{0,v}w_{,r}(0,v)+(1+k_0)M_{1,v}w(0,v)
\end{eqnarray}
where we have written $M=M_0(v)+M_1(v)r+M_2(v)r^2+...$
and $w(r,v)=w(0,v)+w_{,r}(0,v)r+...$.

The first equation can be solved for $M_{0}(v)$
once $k_{0}(v)$ is specified and gives,
\begin{equation}
    M_{0}(v)= C_{1}e^{-3\int_v^1{\frac{k_{0}}{v}dv}}
\end{equation}

From regularity we see that $M_1=0$, and we can
prove also that $w(r,v)$ must go like $r$ near the center,
thus imposing $w(0,v)=0$. From the second equation
we then get $k_1(v)=0$.

In the third equation we use the freedom to
choose $k_{2}(v)$ and take a specific class where
\begin{equation}
    k_2=(1+k_0)k_0\frac{w_{,r}(0,v)}{v}
\end{equation}
This choice allows us to integrate the third equation to obtain
\begin{equation}
    M_{2}(v)=\tilde{C}_{2}e^{-5 \int_v^1{\frac{k_{0}}{v}dv}}
\end{equation}

We have therefore solved the system up to second
order in $r$ and $M(r,v)$ results as,
\begin{equation}\label{mass}
    M(r,v)= C_{1}e^{-3\Phi(v)}\left[1+ r^2 C_{2}e^{-2\Phi(v)}\right]
\end{equation}
where we have defined $C_2=\frac{\tilde{C}_2}{C_1}$ and
\begin{equation}
    \Phi(v)=\int_v^1{\frac{k_0(v)}{v}dv}
\end{equation}

Since we are interested in the behavior near
the center, we have not considered here higher order terms.
Nevertheless we shall notice that the freedom to choose
the equation of state in the form of $k$ allows us to
potentially evaluate all other terms.

We see that positivity of $C_1$ is enough
to ensure positivity of $M$ near the center, while $C_2<0$ would
imply that the density is decreasing radially outwards,
thus imposing certain conditions on the boundary of the
cloud once $C_2$ is set.

The set of Einstein equations is then solved
once we integrate the equation of motion \eqref{mot}.
Since in general it is not possible to integrate
this equation explicitly, we shall restrict ourselves
to a close neighborhood of the center $r=0$ where we can
write the inverse of equation \eqref{mot} and
integrate it to obtain $t(r,v)$ near the center as
\begin{equation}\label{t}
    t(r,v)=t(0,v)+\chi_1(v)r+\chi_2(v)r^2+...
\end{equation}
where the coefficients $\chi_i(v)$ are obtained
from the expansion of $t$ (which is always possible
since the functions involved are generally
at least $\mathcal{C}^2$). These quantities are
related to the structure of the apparent horizon
in the spacetime and the visibility of the singularity
is determined by the same.
The singularity curve is then given
by $t_s(r)=t(r,0)$ and describes the time at
which the shell labeled by the comoving
radius $r$ becomes singular.
In practice it can be shown that positivity
of the first non-vanishing term $\chi_i(0)$ is a
necessary and sufficient condition for the outgoing geodesics
to come out of the singularity \cite{ndim}.

In the present model, it is easy to check that
regularity and the differential equations for $M$
impose $\chi_1(0)=0$.
Therefore, the nature of the singularity is decided
by the sign of $\chi_2(0)$, which is itself
related to the energy density and pressures through $M$.
Since $M$ is given by \eqref{mass} up to
second order, and since we have the freedom to choose
$k_i(v)$ for $i>2$, it is not difficult to show that there
are configurations that lead to the formation of
a timelike naked singularity. Further, it can be shown
that these
configurations are related to the more physically
realistic density profiles where $\rho$ decreases
as $r$ increases (as obtained by choosing $C_2<0$).
The singularity curve turns out to be increasing in
time as $r$ increases and the central shell is the
first to become singular, as would be reasonable
in a physically realistic scenario.

For the sake of clarity, let us briefly summarize
the whole procedure followed above. Einstein's equations,
together with the definition of the Misner-Sharp mass
and the equation of state, constitute a set of 6 equations
in 7 unknown (in $(r,v)$ coordinates they are
$\rho, p, k, t, M, G, H$), leaving us with the freedom
to choose one function at will.
Einstein's equations \eqref{p}, \eqref{rho},
\eqref{G} and \eqref{H} can be integrated to give
$p, \rho, G$ and $H$. The system then reduces to solving
two differential equations for $M(r,v)$ and $t(r,v)$,
coming from equation \eqref{eos} and \eqref{F}, namely
equation \eqref{diff} and the inverse of equation \eqref{mot}.
Restricting ourselves to a close neighborhood of the
center we can expand all the quantities with respect to $r$.
Equation \eqref{diff} decouples from equation \eqref{mot}
for a specific choice of $k$, in which the coefficient
$k_2(v)$ must be fixed. This choice allows us to integrate
equation \eqref{diff}, thus providing the explicit
form of $M(r,v)$ near the center. Using $M$ as obtained
in this manner, together with $\rho, p, G$ and $H$ as obtained
from the other Einstein equations, we can integrate
\eqref{mot} for small values of $r$ to obtain $t(r,v)$ as
in equation \eqref{t}.
Therefore, integrating equation \eqref{mot} in a close
neighborhood of the center and writing $t(r,v)$ as in
equation \eqref{t} solves completely the system of Einstein
equations for small values of $r$, thus providing the space-time
metric of the collapsing cloud approaching the singularity.
The solution retains the freedom to choose the
coefficient $k_0(v)$ from the expansion of
the free function $k(r,v)$.

In order to evaluate the nature of the region
surrounding the singularity curve we shall look for the behavior
of the apparent horizon which determines the boundary
of the trapped surfaces that can eventually form during collapse.
The apparent horizon equation is given by
\begin{equation}\label{ah}
    \frac{F}{R}=\frac{r^2M(r,v)}{v}=1
\end{equation}
which gives implicitly the curve of the apparent horizon as $r_{ah}(v)$.

Typically, in a pressure-free dust collapse, the mass
is conserved and the apparent horizon must have
radius $r_{ah}=0$ at the time of formation of the singularity
(when $v=0$). In that case, the singularity curve for
all $r$ with $r\neq 0$ is entirely trapped, and the central
shell is the only one where the occurrence of the
singularity can be eventually visible.

Nevertheless, in a perfect fluid collapse a wider
variety of options is present. In the case when the mass is entirely
radiated away during collapse as is the case above for the model
we constructed here, we thus have $F(r,0)=0$ for $r>0$,
that is, for a finite range of comoving coordinate values
of $r$.
We thus see from equation \eqref{ah} that no trapped
surfaces form at all, all the way till the singularity
formation for values $r>0$, thus leaving a whole finite
portion of the singularity curve to be visible and timelike.
The resulting naked singularity is timelike with
a range from $r=0$ to a finite value of $r$.
The key point is, when pressures are allowed to be negative,
a finite portion of the non-central singularity at $r>0$
can be visible as shown here.

As stated above, the condition in order to have
the naked singularity to be timelike, is that $F$ or
equivalently $M$
vanishes at $v=0$. In this case we will have
$\frac{F}{R}<1$ in the limit of $v \rightarrow 0$ with
$r \neq 0$. Then the
region surrounding the singularity will not be trapped.
This immediately implies that $M_{,v}>0$
and therefore the pressure, which is given by
$p=-\frac{M_{,v}}{v^2}$, as seen from equation \eqref{p},
must be negative in this case.
Since we have imposed an equation of state of the
kind given by equation \eqref{eos}, negative pressures
imply that in a neighborhood of the singularity
we must have $ k< 0$.

It is indeed possible, and plausible, that
the gas cloud commences collapse with an equation of
state describing
classical positive pressures, and at a later stage, when
the cloud approaches the singularity,
the pressures become negative, probably triggered by
the quantum effects. This finally leads to the evaporation of the core
of the collapsing cloud and to the emission of highly energetic particles.

The free function $k_0$ describes at the lowest order
in $r$ the relation between the pressure and the density.
Namely, it relates the central pressure and the central density
at any given time, and its explicit behavior must
be decided by the physical considerations and this will generally
depend on the nature of the collapsing gas.
Several scenarios where the mass $F$ is radiated away during
collapse can then be devised by choosing suitably the
function $k(r,v)$, and we have seen that quantum effects
could justify $k$ to become negative at some stage
before the formation of the singularity.

As an explicit example, we shall consider $k_0(v)$
to be written as in a series near the singularity
\begin{equation}
    k_{0}(v)=k_{00}+k_{01}v+k_{02}v^2+...
\end{equation}
For simplicity we shall further assume that all higher order terms vanish.
With the choice for $k_{0}$ made above we obtain,
\begin{eqnarray}
  \frac{F}{R}&=&r^2C_{1} v^{-3k_{00}-1} e^{-3k_{01}v-\frac{3}{2}k_{02}v^2}\cdot \\ \nonumber
  && \cdot\left[ 1+r^2C_{2} v^{-2k_{00}} e^{-2k_{01}v-\frac{2}{2}k_{02}v^2}\right]
\end{eqnarray}

The requirement that $\frac{F}{R}<1$ near the
singularity imposes the following restriction on $k_{00}$,
\begin{equation}
    k_{00} \leq -\frac{1}{3}
\end{equation}
which is a condition for avoidance of trapped surfaces,
and which gives rise to the formation of a non-central
naked singularity.
We can choose $k_{01}$ and $k_{02}$ so that the
pressure is positive to begin with, at the onset of collapse
for large values of $v$ and at the initial time.
The pressure then decreases during collapse and turns
negative at the later stages of collapse,
eventually leading to the formation of a naked singularity.

We can see that in the limiting case where
$k_{00}=-\frac{1}{3}$, then in the limit of approach of
singularity we get,
\begin{equation}
    \lim_{v\rightarrow 0}\frac{F}{R}= r^2 C_{1}
\end{equation}
which implies a finite radius for the trapped surfaces
at all times. This in turn means that we must take the
boundary of the collapsing cloud according to
\begin{equation}
    r_{b}<\frac{1}{\sqrt{C_{1}}}
\end{equation}
in order to avoid trapped surfaces during
the whole collapse.
In the case where $k_{00}<-\frac{1}{3}$, the
apparent horizon simply does not form as the singularity is
approached, thus suggesting that the boundary chosen
in order to avoid trapped surfaces can be
arbitrarily large in this case.

As we have said, the presence of negative pressure
requires that the function $M$ vanishes at the singularity,
which implies that the whole mass of the core must be radiated
away during the last stages of collapse. This shows how
in these collapse models a finite amount of energy escapes
away from the vicinity of the naked singularity.
In the classical picture, this mechanism requires the
presence of negative pressures, which can be justified by
quantum effects occurring near the singularity, as it
has already been shown both in the semiclassical approximation
as well as in some simple toy models in Loop Quantum Gravity
\cite{Singh}.
Particles escaping the vicinity
of the singularity by this mechanism would have
to be intrinsically highly energetic since they originate in the
region where quantum gravity is dominant. These particles,
when traveling close to the Cauchy horizon, would retain their
energy even at large distances from the singularity
as they travel in the region of spacetime which can be described by a
metric which is Minkowski with small perturbations.

In this sense, such a timelike naked singularity can
provide a window on the physics of Planck scale.
These particles would travel in a region close to the
Cauchy horizon and participate in the large center of
mass energy collisions with ingoing particles,
as was described in section II.

\section{Conclusion}

We showed here that naked singularities that form in
gravitational collapse of massive stars, evolving from a regular
initial data, can act as cosmic super-colliders. Particles
interact and collide near the Cauchy horizon of the naked
singularities with arbitrarily high center of mass energies.
This might provide us the natural laboratories to study
the physics of collisions with center of mass energies significantly large,
as compared to those realized in terrestrial particle accelerators,
and these could go all the way high, up to the Planck scale
in an astrophysical setting. This scenario is subject to the existence
of naked singularities in nature, as produced in astrophysical
events such as collapse of massive stars.

Given the existence of many solutions to Einstein
equations admitting naked singularities in general relativistic
gravitational collapse of physically realistic matter fields,
it might be reasonable to expect that naked singularities
might occur in nature, and if they do so, then they would
turn out to be a boon for our understanding of various physical
processes occurring at ultra-high energies, not realizable
in any other laboratory or astrophysical settings, apart
from perhaps the very early universe. Naked singularities
would accelerate particles to arbitrarily high energies,
which travel close to the Cauchy horizon, and these particles
would participate in high energy collisions. We also considered
and discussed here various physical mechanisms responsible
for production of such particles by naked singularities.
Incoming particles falling inwards from infinity just missing
the singularity, and then following the outward trajectory,
would be extremely energetic. They perhaps might get an additional
boost due to the repulsive nature of naked singularities at
classical and quantum level. They would travel close to the
Cauchy horizon and collide with incoming particles at large
center of mass energies.

During the final stages of gravitational collapse, the
density and curvatures reach the Planck scale, where dynamics
of collapse is ruled by quantum gravity. Quantum gravity
is known to be ultra-repulsive and is expected to resolve
the classical naked singularity. Calculations carried out in
semiclassical regime strongly indicate that there would be
a strong outburst of extremely large amount of energy and
central fluid elements might be thrown apart at gigantic
energies in full quantum theory of gravity, when the discrete
evolution is taking into account. When the ultra-relativistic
fluid particles bounce back and travel close to what would have
been a Cauchy horizon in classical picture, they collide with
fluid particles from the outer shells undergoing collapse.
Thus there would be large number of collisions with arbitrarily
large energies, followed by more collisions between energetic
particles emitted from the primary collisions and the outer
fluid elements in the collapsing cloud. This can generate
a quantum fireball, physical processes occurring in which would
be governed by quantum gravity.

We also proposed another situation
where collapse of the central shell would yield momentary
singularity but the noncentral shell, because of large inhomogeneity
in pressure, would be blown apart to ever increasing velocity,
eventually reaching asymptotically to the Cauchy horizon of
the singularity, where collision with ingoing particles would
reach large center of mass energies.
Thus, we demonstrated that the naked singularity, if it were
to occur in nature, could turn out to be a boon for understanding
phenomenon at and around the Planck scale.

We have described here a model to demonstrate how
the repulsive nature of gravity, as modeled by
large negative pressure in the vicinity of naked singularity,
accounts for an outward flux of energy in the form of highly
energetic photons or massive particles.
We studied the final stages of a collapsing cloud
consisting of an ideal fluid with a variable equation of state.
We chose an equation of state so that the pressure becomes negative
in the vicinity of the naked singularity.
This phenomenon is well motivated from the quantum gravity
effects near the singularity.
The existence of negative pressure essentially implies
the decrease in the Misner-Sharp mass of the cloud, in the
form of an outward flux of
energy which can be modeled by exterior generalized
Vaidya spacetime. If the mass is radiated sufficiently fast then
the formation of trapped surfaces can be avoided,
thus exposing the singularity. The entire mass of the cloud
can also be radiated away, leaving behind a flat spacetime.
There is an outgoing radiation from close to
naked singularity which consists of highly energetic
photons or massive particles, since it originates in the
region with extremely high curvatures dominated by quantum gravity.
When these particles collide and interact with ingoing
particles, the center of mass energy of collision would be extremely
large, offering a window to new physics at energies
inaccessible at terrestrial particle accelerators.

We thank R. Goswami, F. de Felice, and K. Nakao
for comments and discussions.

\end{document}